\documentclass[pdflatex,sn-basic]{sn-jnl}
\usepackage{graphicx}%
\usepackage{multirow}%
\usepackage{amsmath,amssymb,amsfonts}%
\usepackage{amsthm}%
\usepackage{mathrsfs}%
\usepackage[title]{appendix}%
\usepackage{xcolor}%
\usepackage{textcomp}%
\usepackage{manyfoot}%
\usepackage{booktabs}%
\usepackage{algorithm}%
\usepackage{algorithmicx}%
\usepackage{algpseudocode}%
\usepackage{listings}%

\theoremstyle{thmstyleone}%
\theoremstyle{thmstyletwo}%

\theoremstyle{thmstylethree}%

\begin{document}

\title{Probing Materials Knowledge in LLMs: From Latent Embeddings to Reliable Predictions}

\author*[1]{\fnm{Vineeth} \sur{Venugopal}}\email{vineethv@mit.edu}
\author[1]{\fnm{Soroush} \sur{Mahjoubi}}
\author[1]{\fnm{Elsa} \sur{Olivetti}}

\affil*[1]{\orgdiv{Department of Materials Science and Engineering}, \orgname{Massachusetts Institute of Technology}, \orgaddress{\city{Cambridge}, \postcode{02139}, \state{MA}, \country{USA}}}

\abstract{
Large language models are increasingly applied to materials science, yet fundamental questions remain about their reliability and knowledge encoding. Evaluating 25 LLMs across four materials science tasks---over 200 base and fine-tuned configurations---we find that output modality fundamentally determines model behavior. For symbolic tasks, fine-tuning converges to consistent, verifiable answers with reduced response entropy, while for numerical tasks, fine-tuning improves prediction accuracy but models remain inconsistent across repeated inference runs, limiting their reliability as quantitative predictors. For numerical regression, we find that better performance can be obtained by extracting embeddings directly from intermediate transformer layers than from model text output, revealing an ``LLM head bottleneck,'' though this effect is property- and dataset-dependent. Finally, we present a longitudinal study of GPT model performance in materials science, tracking four models over 18 months and observing 9--43\% performance variation that poses reproducibility challenges for scientific applications.
}


\maketitle


\section{Introduction}\label{sec:intro}

Large language models represent a fundamentally different paradigm for materials science compared to traditional machine learning approaches. Conventional models---whether random forests, support vector machines, or graph neural networks---require explicit featurization: transforming chemical compositions into numerical descriptors such as elemental properties, structural fingerprints, or graph representations~\cite{dunn2020benchmarking}. This featurization step demands domain expertise, constrains model inputs to predefined formats, and limits the types of queries that can be posed. LLMs operate differently, offering several capabilities that distinguish them from traditional approaches:

\begin{itemize}
\item \textbf{Natural language input}: LLMs accept prompts directly (``predict the bandgap of BaTiO$_3$''), eliminating the need for explicit feature engineering and enabling intuitive interaction with materials data~\cite{jablonka2024llm}.
\item \textbf{Robustness to input variations}: LLMs recognize the equivalence between ``barium titanate,'' ``BaTiO$_3$,'' and ``TiBaO$_3$'' without explicit mapping rules~\cite{polak2024chatextract}.
\item \textbf{Multi-step workflow orchestration}: Through agent frameworks and tool integration, LLMs can retrieve literature, query databases, run simulations, and synthesize results into coherent analyses~\cite{bran2024chemcrow, ramos2025review}.
\item \textbf{Rich textual responses}: Unlike models that output single numerical predictions, LLMs generate explanations, caveats, and reasoning chains alongside their answers.
\item \textbf{Emergent capabilities}: LLMs can perform tasks for which they were not explicitly trained, exhibiting zero-shot transfer that traditional task-specific models cannot match~\cite{brown2020gpt3}.
\end{itemize}

\noindent These features have driven rapid adoption of LLMs across materials science. In literature mining, transformer-based models now extract structured data from unstructured text at scale—MaterialsBERT has processed over 300,000 property records from scientific abstracts~\cite{shetty2022materialsbert}, while named entity recognition systems identify materials, properties, and processing conditions for millions of papers~\cite{weston2019named, schilling2025text}. For property prediction, fine-tuned LLMs have achieved competitive or superior performance compared to graph neural networks and traditional machine learning approaches. LLM-Prop demonstrates 8--65\% improvement over GNNs for crystal property prediction~\cite{rubungo2025llmprop}, while ElaTBot reduces elastic tensor prediction errors by 33\% compared to domain-specific models~\cite{liu2025elatbot}. Beyond prediction, LLMs now generate novel crystal structures directly from textual descriptions~\cite{antunes2024crystallm}, plan synthesis routes~\cite{kim2020inorganic}, and orchestrate autonomous experimental workflows through agent frameworks that integrate literature retrieval, database queries, and simulation tools~\cite{bran2024chemcrow, kang2024chatmof, macknight2025rethinking}.

However, basic questions about LLM behavior in materials applications remain poorly understood~\cite{zaki2024mascqa, mirza2025chembench, miret2025matsci, wang2025robustness, chiang2024llamp}. How is information encoded in these models and does this encoding vary by task or property? What determines whether fine-tuning succeeds or fails? Do these models develop genuine understanding of material entities, or do they merely learn statistical patterns? Does model size matter? Can fine-tuning compensate for differences in pre-training performance? Are there differences in performance between open models and those served through APIs? How does hallucination affect task performance and can hallucination be mitigated through fine-tuning?

This work addresses these questions through a comprehensive evaluation of 25 large language models across four materials science tasks: bandgap prediction, dielectric constant prediction, crystal system classification, and knowledge graph completion — spanning both symbolic and numerical output modalities. We assess more than 200 base and fine-tuned model configurations, making this one of the largest such evaluations in materials science to date.

Our central finding is that output modality fundamentally determines LLM behavior. We discover a striking asymmetry in how models fail and improve: for symbolic tasks, base models exhibit high output variability — termed `response entropy' — coupled with low accuracy, reflecting a knowledge gap that fine-tuning effectively closes to produce convergent, correct outputs. For numerical tasks, base models instead show confident hallucination — low entropy despite poor accuracy — and fine-tuning improves predictions without resolving this false confidence, urging caution when interpreting model certainty.

Investigating the mechanisms underlying each modality, we find that symbolic task performance is driven by how frequently an entity appears across diverse training contexts — indicating that fine-tuning builds distributional representations. For numerical tasks, layer-wise embedding probes reveal an ``LLM head bottleneck'': intermediate representations encode more predictive information than models can express through text generation, though this effect is property-dependent — prominent for bandgap but diminished for dielectric constant. Cross-task transfer analysis further indicates that fine-tuning primarily confers domain adaptation rather than modality-specific skills.

Finally, we present an 18-month longitudinal study of GPT models revealing 9–43\% performance variation, posing significant challenges for reproducible materials science research. Together, these findings provide practical guidance for deploying LLMs in materials informatics while revealing fundamental limitations in how these models represent and express scientific knowledge.


\section{Results}\label{sec:results}

25 large language models are evaluated across four materials science tasks that span two distinct output modalities. The symbolic tasks are ontological reasoning, where knowledge graph triples from MatKG are recast as natural language question-answer pairs (e.g., the triple (BaTiO\textsubscript{3}, has\_property, high\_dielectric\_constant) becomes ``What are the key properties of BaTiO\textsubscript{3}?''), and crystal system classification, where models assign compounds to one of seven crystal systems from their chemical formula. The numerical tasks are bandgap prediction and dielectric constant prediction, where models generate continuous property values from chemical composition alone. While composition is not a sufficient descriptor for these properties — which also depend on crystal structure, processing, and microstructure — our goal is not to achieve state-of-the-art predictive accuracy but rather to study how LLMs encode and express numerical material knowledge. This task selection enables direct comparison of LLM behavior across output modalities while covering both structure-property relationships and knowledge retrieval. We assess both base and fine-tuned performance, with response entropy computed across 10 independent inference runs to quantify output consistency. Full dataset and training details are provided in Methods.

Across all four tasks, fine-tuning yields substantial improvements over base model performance, with detailed analyses presented in the task-specific sections below. Table~\ref{tab:summary} provides a bird's-eye view, summarizing the top-performing fine-tuned models across all four tasks. GPT-4.1-mini ranks first on three of four tasks, making it the most consistently strong performer, while open-weights models such as Llama-3-70B match or exceed proprietary systems on specific tasks. However, these headline numbers mask a fundamental divergence in how LLMs behave across output modalities, which we examine in the following sections.

\begin{table}[ht]
\centering
\caption{\textbf{Top three fine-tuned models per task.} RMSE (lower is better) for regression; accuracy (higher is better) for symbolic tasks. See Methods for metric definitions and Supplementary Tables S5.1--S5.4 for all 25 models.}
\label{tab:summary}
\begin{tabular*}{\textwidth}{@{\extracolsep{\fill}}cllll@{}}
\toprule
 & Bandgap & Dielectric & Crystal System & MatKG \\
 & (RMSE, eV) & (RMSE) & (Acc., \%) & (Top-1, \%) \\
\midrule
1 & Llama-3-70B (0.70) & GPT-4o (1.8) & GPT-4.1-mini (66) & GPT-4.1-mini (68) \\
2 & GPT-4.1-mini (0.71) & GPT-4.1-mini (1.9) & Llama-3-70B (63) & GPT-4.1 (66) \\
3 & GPT-4o-mini (0.73) & Llama-3-70B (2.1) & GPT-4o-mini (63) & GPT-4o (65) \\
\bottomrule
\end{tabular*}
\end{table}
\subsection{Symbolic tasks: Link prediction and classification}

\begin{figure}[ht]
\centering
\includegraphics[width=\textwidth]{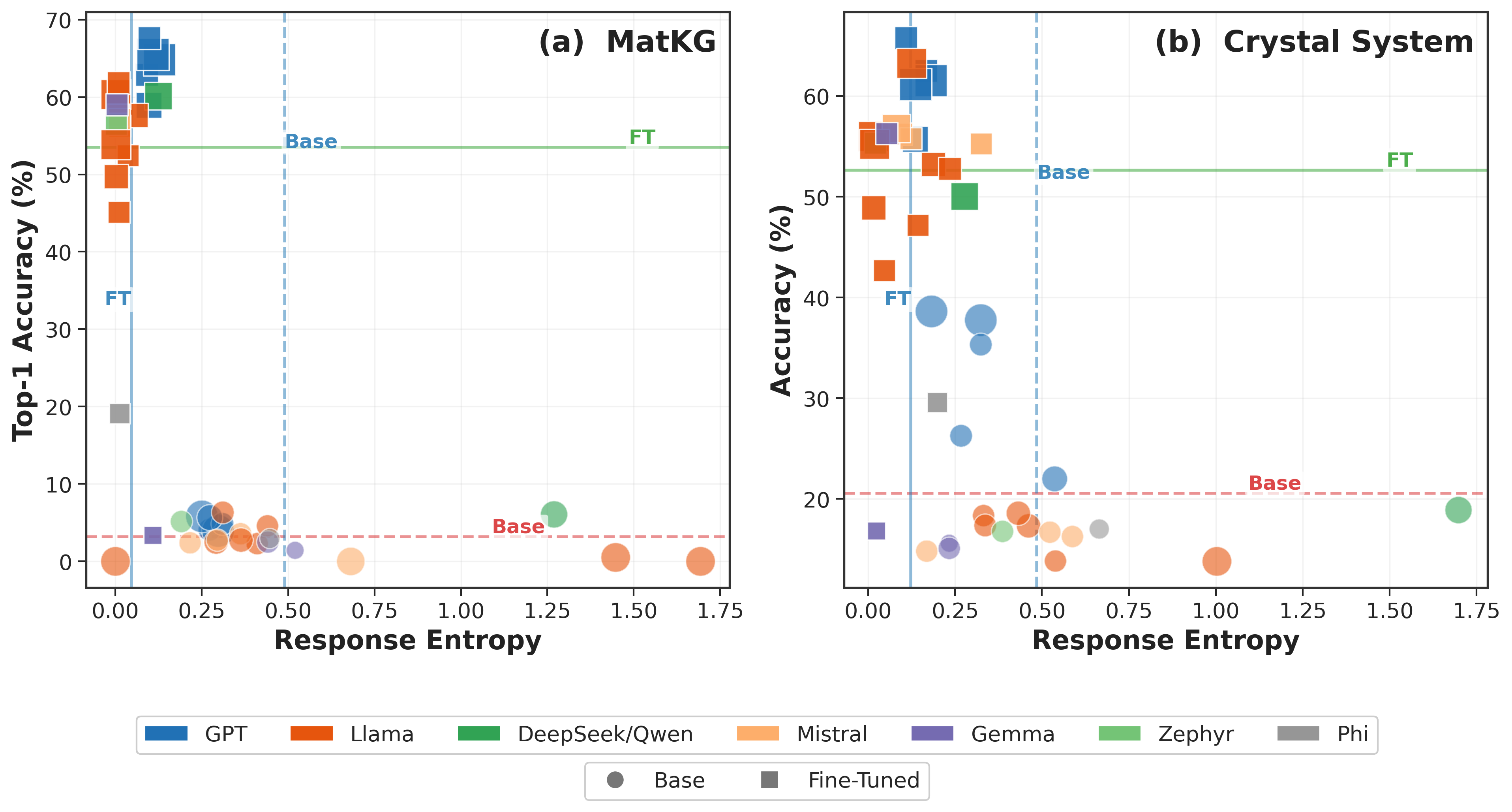}
\caption{\textbf{Performance versus response entropy for symbolic tasks.} MatKG link prediction (left) and crystal system classification (right). Circles indicate base models; squares indicate fine-tuned models. Marker size is proportional to model parameter count, and colors denote model families. Dashed lines show mean performance and entropy for base models. Fine-tuning dramatically improves accuracy while reducing response entropy across all model families.}
\label{fig:classification}
\end{figure}

Base models perform poorly on both symbolic tasks regardless of architecture or scale (Figure~\ref{fig:classification}). MatKG link prediction yields only 2--6\% Top-1 accuracy, barely above random chance, while crystal system classification reaches 13--39\% across seven classes. Both tasks show high response entropy coupled with low accuracy, indicating that models fail to converge on consistent outputs across runs.

Fine-tuning improves the performance of these symbolic tasks, consistent with recent findings across materials and chemistry domains~\cite{vanherck2025assessment, zhang2024finetuning}. For MatKG, GPT-4.1-mini achieves the highest fine-tuned accuracy at 68\% (+63pp), followed closely by GPT-4.1 (66\%) and GPT-4o (65\%). Among open-weight models, Llama-3-8B reaches 62\% and Llama-2-70B achieves 60\%, demonstrating that mid-sized open models can approach proprietary model performance after task-specific training. Crystal system classification follows similar patterns: GPT-4.1-mini leads at 66\%, with Llama-3-70B (63\%) and GPT-4o (62\%) close behind. The GPT family maintains an advantage, but the gap narrows from 20+ percentage points at baseline to approximately 5--10 points after fine-tuning.

Entropy changes mirror the accuracy improvements. Fine-tuning reduces response entropy by 25--99\% across models, with the largest reductions observed in models achieving the highest accuracies. For MatKG, Zephyr-7B shows a 99\% entropy reduction while reaching 57\% accuracy; GPT-4.1-mini reduces entropy by 68\% en route to 68\% accuracy. This pattern--- reduced entropy accompanying improved accuracy---indicates that fine-tuning teaches models to converge consistently to verified outputs rather than merely shifting the distribution of guesses. Full results are provided in Supplementary Tables S5.3 and S5.4.

\subsection{Numerical regression tasks}

\begin{figure}[ht]
\centering
\includegraphics[width=\textwidth]{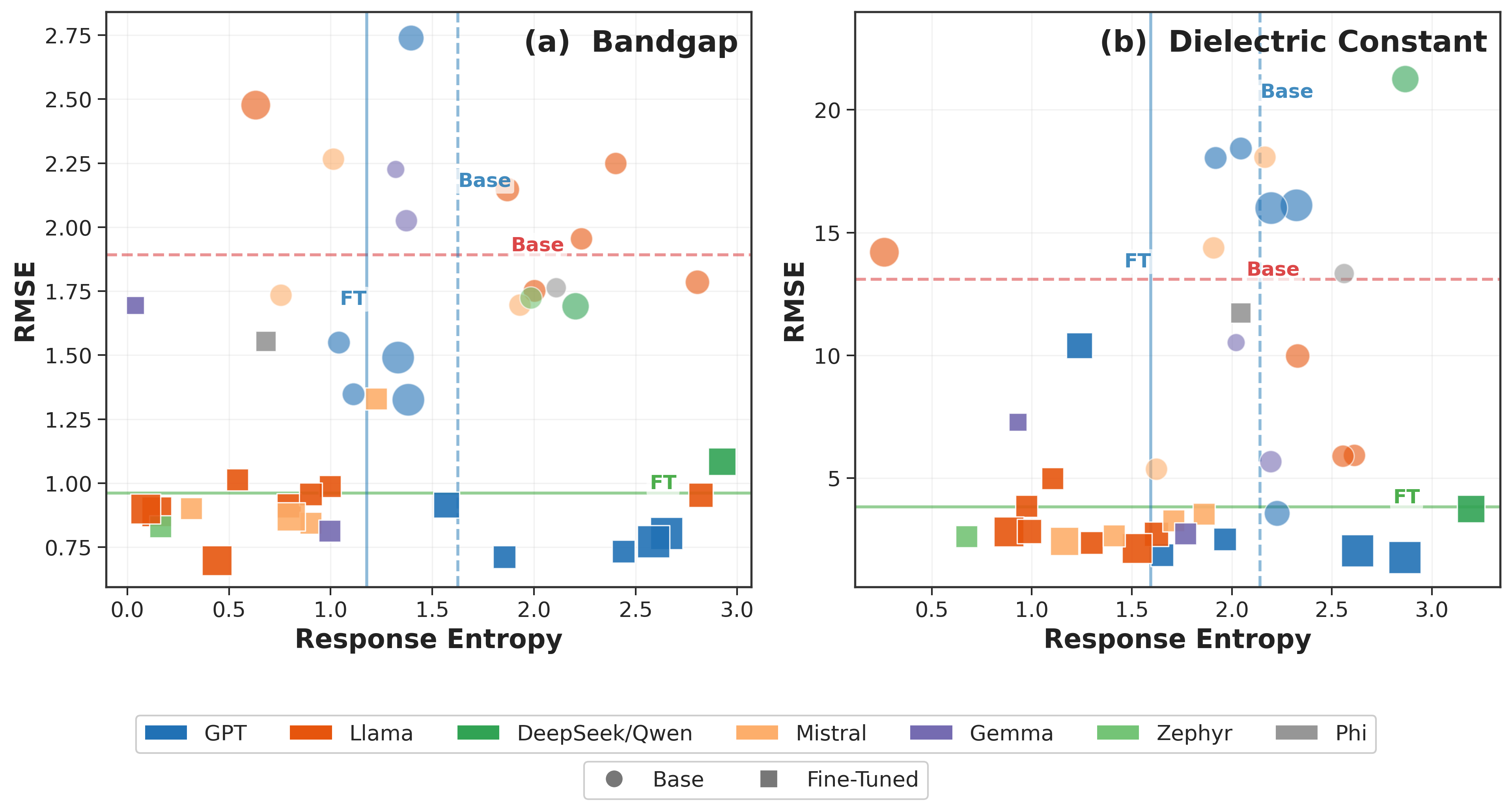}
\caption{\textbf{Performance versus response entropy for numerical regression tasks.} Bandgap prediction (left) and dielectric constant prediction (right). Visualization follows the same conventions as Figure~\ref{fig:classification}. Lower RMSE indicates better performance. Fine-tuning generally reduces prediction error, with variable effects on entropy across models.}
\label{fig:regression}
\end{figure}

Numerical property prediction---bandgap and dielectric constant---is evaluated using root mean square error (RMSE), with response entropy computed from the variance across 10 inference runs (Figure~\ref{fig:regression}). Unlike symbolic tasks, base models exhibit low entropy despite poor performance---a "confident hallucination". Models generate precise-looking numerical outputs with high consistency across runs, yet these outputs bear little relationship to ground truth. Base model RMSE varies widely: GPT-4.1 achieves 1.3 eV for bandgap while Llama-2-13B reaches 2.2 eV; for dielectric constant, the spread is even larger (5.4 to 29 RMSE).

Fine-tuning substantially improves numerical prediction. For bandgap, Llama-3-70B achieves the best fine-tuned RMSE of 0.70 eV (+72\% improvement), followed by GPT-4.1-mini (0.71 eV, +47\%) and GPT-4o-mini (0.73 eV, +53\%). The GPT family and open-weights models perform comparably after fine-tuning: GPT-4o reaches 0.80 eV while Llama-3-8B achieves 0.96 eV and Mistral-7B-v0.1 reaches 0.90 eV. Dielectric constant shows even larger relative improvements: GPT-4o achieves the lowest fine-tuned RMSE of 1.8 (+89\%), with GPT-4.1-mini (1.9, +90\%) and Llama-3-8B (2.4, +91\%) following. Open-weights models again approach GPT performance levels, with Llama-2-13B improving from 29 to 2.7 RMSE (+91\%).

Entropy behavior differs markedly from symbolic tasks. Fine-tuning produces variable entropy changes for numerical prediction: some models show entropy increases (GPT-4o bandgap: +99\%; Llama-3-70B dielectric: +480\%), while others show decreases (Zephyr-7B bandgap: $-$92\%; Llama-3-8B dielectric: $-$50\%). This inconsistency suggests that for numerical tasks, entropy reflects output format consistency rather than prediction confidence---a model may become more variable in its numerical formatting while improving accuracy, or vice versa. Full results are provided in Supplementary Tables S5.1 and S5.2.

\subsection{Output modality determines performance characteristics}

A central finding of this work is that LLM performance differ fundamentally between output modalities. Symbolic tasks (classification, link prediction) show uniformly poor base performance across all models, with fine-tuning providing dramatic improvements of 30--60 percentage points. In contrast, numerical regression tasks exhibit high variance in base performance, with some models achieving reasonable predictions without fine-tuning, and improvement magnitudes that depend strongly on the specific model-task combination.

This modality dependence extends to the relationship between model scale and performance. For symbolic tasks, larger models generally perform better both before and after fine-tuning, with a clear scaling trend. For numerical tasks, this relationship is less consistent: GPT-4.1 achieves the best base bandgap RMSE, but the 8B-parameter Llama-3 matches or exceeds many larger models after fine-tuning. Complete results for all model-task combinations are provided in Supplementary Tables S1--S4.

\subsection{Cross-task transfer}

\begin{figure}[ht]
\centering
\includegraphics[width=0.9\textwidth]{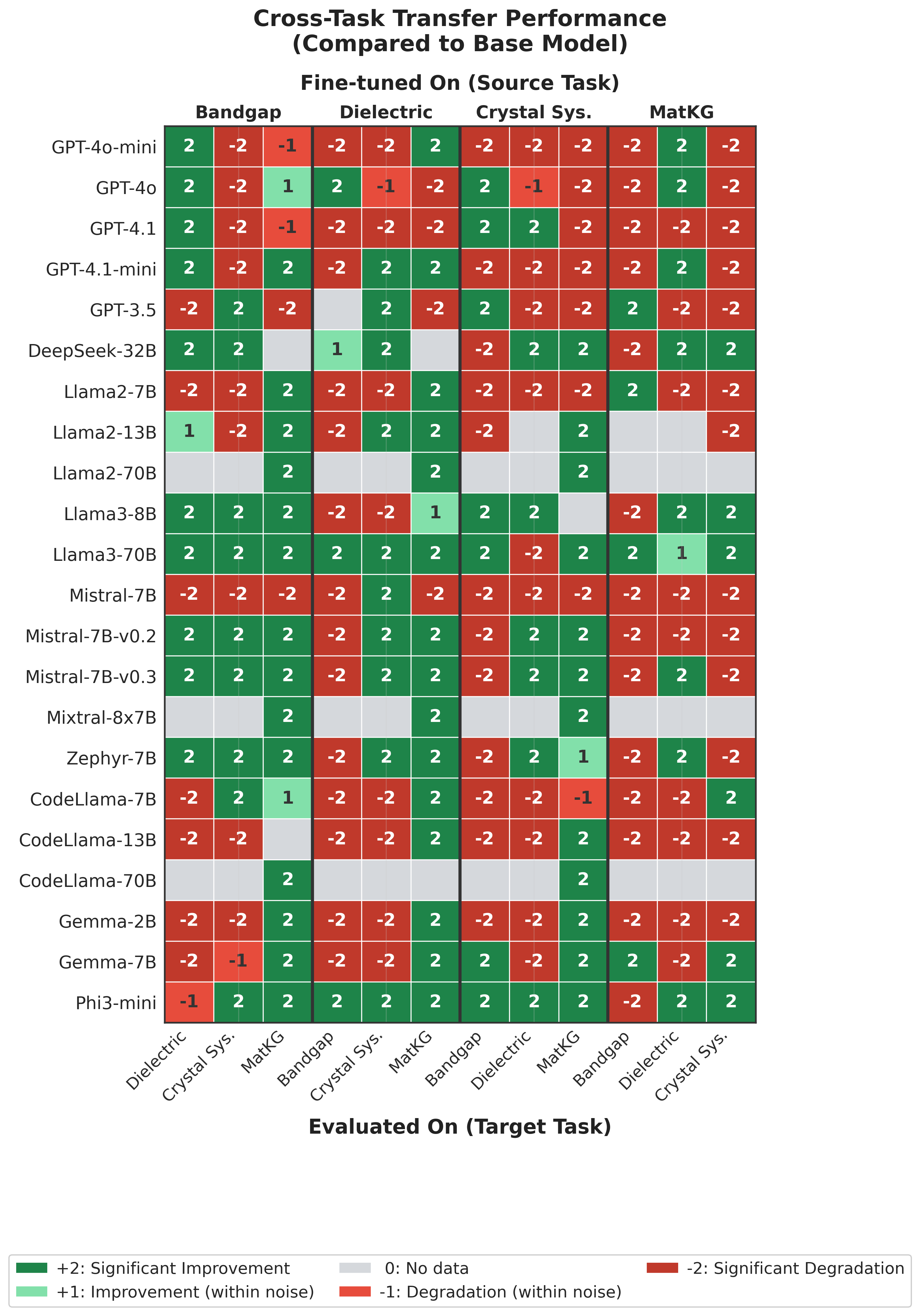}
\caption{\textbf{Cross-task transfer matrix.} Each cell shows the effect of fine-tuning on a source task (columns) when evaluated on a target task (rows), compared to base model performance. Color coding: dark green (+2) indicates statistically significant improvement; light green (+1) indicates improvement within noise; light red (--1) indicates degradation within noise; dark red (--2) indicates significant degradation; grey (0) indicates no data. Statistical significance is determined by comparison to base model standard deviation across inference runs.}
\label{fig:transfer}
\end{figure}

To investigate whether materials knowledge transfers across tasks, we evaluated each fine-tuned model on all four tasks and compared performance to base models (Figure~\ref{fig:transfer}). Statistical significance was determined by comparing performance changes to base model variance: differences exceeding one standard deviation were marked as significant. Full transfer matrices for all models and details on the statistical significance methodology are provided in Supplementary Sections S6 and S6.23.

The strongest positive transfer occurs between the two numerical regression tasks. Models fine-tuned on bandgap often improve dielectric prediction: GPT-4.1 achieves 7.9 RMSE on dielectric when fine-tuned on bandgap (vs. 16 base), and Llama-3-70B reaches 10 (vs. 14 base). The reverse transfer is also effective---dielectric fine-tuning improves bandgap prediction for GPT-4o (1.4 eV vs. 1.5 base) and Mixtral-8x7B (1.3 eV vs. 1.7 base). This bidirectional transfer suggests that numerical regression tasks share composition-property representations that generalize across properties.

Transfer between symbolic and numerical tasks shows consistent failure patterns. Fine-tuning on MatKG or crystal system classification rarely improves numerical prediction beyond base model levels, and numerical fine-tuning provides minimal benefit for symbolic tasks. For instance, models fine-tuned on bandgap achieve only 13--35\% accuracy on crystal system classification, comparable to base model performance (13--39\%). Similarly, MatKG-fine-tuned models show no meaningful improvement on bandgap or dielectric prediction. This asymmetry suggests that symbolic and numerical tasks require distinct representational adaptations that do not necessarily transfer across modalities.

Notably, we observe minimal evidence of catastrophic forgetting---fine-tuning on one task rarely causes significant degradation on unrelated tasks. Most off-diagonal entries in the transfer matrix show changes within the noise range of base model performance, suggesting that fine-tuning preserves general model capabilities while adding task-specific knowledge.

\section{Discussion}\label{sec:discussion}

\subsection{Entropy signatures reveal modality-dependent confidence patterns}

The relationship between response entropy and model performance suggests an asymmetry between symbolic and numerical tasks that, to our knowledge, has not been previously reported. For symbolic tasks (MatKG, crystal system classification), base models exhibit high entropy coupled with low accuracy, indicating lack of token-level convergence across inference runs. Fine-tuning reduces entropy while improving accuracy, indicating that models converge to consistent (and verifiable) outputs. We note that this response entropy measures token-level consistency across repeated inference runs, not uncertainty in the sense used in machine learning for physical systems---it reflects whether the model converges to the same answer, not whether that answer is calibrated to true probabilities. This pattern suggests that symbolic task performance is limited primarily by knowledge rather than expression: base models lack the domain-specific patterns needed for accurate prediction, and fine-tuning instills this knowledge.

Numerical regression tasks present a different entropy landscape. Base models often show low entropy despite poor performance---confident hallucination~\cite{mirza2025chembench, wang2025robustness}. This phenomenon is not unique to materials science: overconfident yet incorrect outputs have been documented in mathematical reasoning~\cite{boye2025reasoning, rahman2025fragile} and chemical question answering~\cite{schilling2025text}, suggesting a general limitation of autoregressive language models rather than a domain-specific artifact. Models generate precise-looking numerical outputs with high consistency across runs, yet these outputs bear little relationship to ground truth values. Fine-tuning reduces RMSE substantially while entropy changes are more modest, suggesting that the challenge lies not in model confidence but in the accuracy of confidently-held predictions. This distinction has important practical implications: for symbolic tasks, high entropy can serve as a reliable indicator of uncertain predictions, enabling appropriate abstention or human review. For numerical tasks, entropy provides weaker guidance---a model may be confidently wrong, and low entropy offers false reassurance about prediction reliability.

The entropy asymmetry points to different failure modes for different output modalities. Symbolic task errors reflect knowledge gaps: without fine-tuning, models lack the vocabulary of materials relationships needed for accurate completion. Numerical task errors reflect verbalization failures: models may encode relevant information internally but struggle to express it as precise numerical values through autoregressive token generation~\cite{arai2025numerical, yang2025robustness}. This interpretation motivates our investigation of layer-wise embeddings---if numerical information is encoded but poorly verbalized, probing internal representations should reveal the hidden knowledge.

A notable finding across both modalities is that model scale has limited impact on fine-tuned performance once a minimum capacity threshold is exceeded. For base models, larger GPT variants show modest advantages (GPT-4.1 achieves 1.3 eV bandgap RMSE vs. 1.6 eV for GPT-4o-mini), but this relationship is inconsistent among open-weights families---Llama-3-70B performs worse than Llama-3-8B on several base tasks. After fine-tuning, however, model size becomes largely irrelevant: GPT-4o-mini (0.73 eV) outperforms GPT-4o (0.80 eV) on bandgap, Llama-3-8B (62\%) approaches Llama-2-70B (60\%) on MatKG, and 7B Mistral variants match or exceed 13B CodeLlama models across tasks. The exceptions---Gemma-2B and Phi-3-mini---show consistently poor fine-tuned performance (1.7 eV bandgap, 17\% crystal system accuracy for Gemma-2B), suggesting a capacity floor around 7B parameters below which models cannot effectively learn materials science tasks through LoRA fine-tuning.

\subsection{What determines knowledge graph completion performance?}

The MatKG link prediction task provides a window into how LLMs represent and retrieve materials science knowledge. MatKG is constructed from a materials science knowledge graph~\cite{venugopal2024matkg} in which relations are symmetric: every triple such as (TiO$_2$, \textit{has-property}, piezoelectric) has a corresponding reverse (piezoelectric, \textit{property-of}, TiO$_2$). However, the link prediction dataset retains only the top 5 objects by co-occurrence frequency for each (Subject, Relation) pair. This truncation breaks the underlying symmetry---``piezoelectric'' may appear as a top-5 property for TiO$_2$, but TiO$_2$ need not rank among the top 5 materials for piezoelectricity, where it competes against hundreds of alternatives for those answer slots. The result is a structural asymmetry between how entities function as query subjects versus answer objects.

\begin{figure}[ht]
\centering
\includegraphics[width=\textwidth]{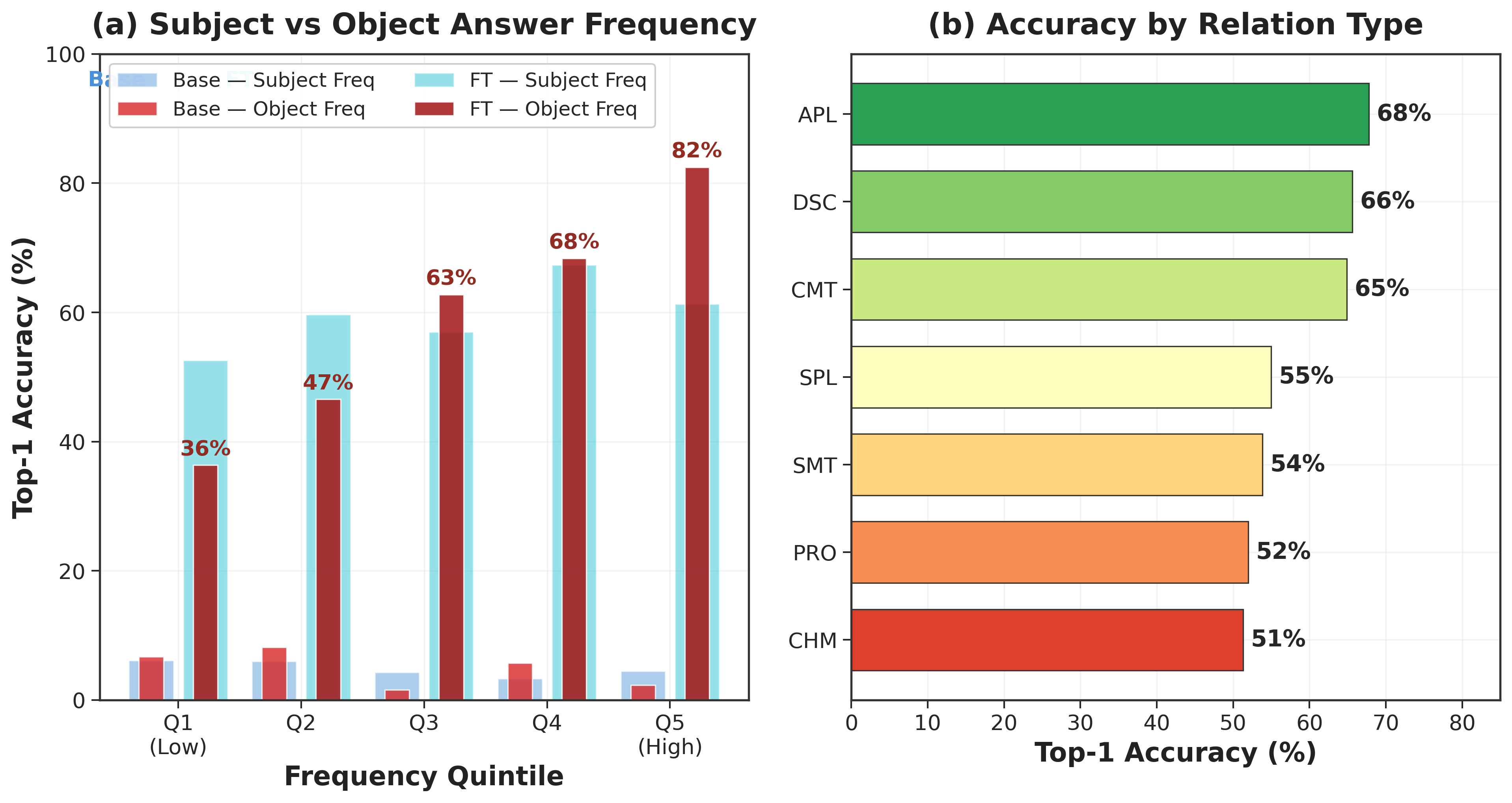}
\caption{\textbf{Factors determining MatKG link prediction accuracy.} (a) Accuracy binned by subject frequency (wide bars) and object answer frequency (narrow bars) for both base and fine-tuned models. The fine-tuned model shows strong monotonic dependence on object answer frequency while subject frequency has no effect; the base model remains flat at $\sim$4--5\% regardless of frequency. (b) Performance breakdown by relation target category, showing that Descriptors (DSC) and Applications (APL) are easiest to predict ($\sim$70\% accuracy) while Symmetry Phase Labels (SPL), Synthesis Methods (SMT) and Materials (CHM) are harder ($\sim$55\%).}
\label{fig:matkg_analysis}
\end{figure}

Figure~\ref{fig:matkg_analysis}(a) illustrates this asymmetry directly: we binned test queries by the subject frequency and object answer frequency of their ground truth, respectively, and plot accuracy for both base and fine-tuned models. The fine-tuned model shows strong monotonic dependence on object answer frequency, while subject frequency has no discernible effect. The base model remains flat at approximately 4--5\% across all quintiles, confirming that the frequency-dependent performance is a product of fine-tuning rather than a prior bias inherited from pretraining. Other graph-structural features---including node centrality and domain specificity---are similarly non-predictive (see S7).

Object answer frequency captures how extensively an entity appears across diverse training contexts, thereby reflecting the richness of its distributional representation acquired during fine-tuning. An entity like PZT, appearing across hundreds of training prompts as an answer to queries about ferroelectricity, perovskite structure, sensor applications, and $d_{33}$ coefficients, develops a dense web of indirect associations with co-occurring concepts. At test time, when the model encounters a novel query about piezoelectric materials---a prompt absent from training---it can retrieve PZT through these shared contextual neighbors: many of the terms the model learned to associate with PZT during training were themselves associated with piezoelectricity in other training prompts, even though the direct PZT--piezoelectric pairing was never observed. This mechanism resembles classical distributional semantics---``you shall know a word by the company it keeps''~\cite{firth1957synopsis}---operating through the fine-tuning gradient rather than through pretraining corpus statistics. A rare entity appearing in only a handful of training prompts lacks this context  and is correspondingly harder to retrieve.

Category-level analysis further supports this interpretation, as shown in Figure~\ref{fig:matkg_analysis}(b). Predictions involving Descriptors (DSC) and Applications (APL) achieve approximately 70\% accuracy, while Symmetry Phase Labels (SPL) and Synthesis Methods (SMT) reach only 55\%. This aligns with the distributional mechanism: descriptive terms and application domains are generic enough to appear as answers across many training prompts, building rich contextual profiles. Crystallographic notation,  synthesis procedures and materials are more specific, appear in fewer contexts, and therefore develop sparser distributional representations~\cite{zaki2024mascqa}.

This is relational learning, but relational in the statistical sense rather than the physical---the model has no representation of \textit{why} PZT is piezoelectric, only that PZT occupies a representational neighborhood densely connected to piezoelectric-adjacent concepts. This aligns with recent work questioning whether LLMs perform genuine reasoning or sophisticated pattern matching~\cite{boye2025reasoning, mirza2025chembench, rahman2025fragile}. The practical consequence is that prediction quality scales with the diversity of training contexts an entity appears in, and current LLMs will reliably complete relationships involving well-attested entities but struggle with rare combinations.

\subsection{Layer-wise probing reveals property-dependent encoding}

While fine-tuning substantially improves numerical regression performance, the best fine-tuned models still achieve RMSE values (0.70--0.85 eV for bandgap) that lag behind state-of-the-art graph neural network approaches~\cite{matbench2020}. This performance gap is consistent with a growing body of literature demonstrating that LLMs struggle with numerical reasoning tasks compared to textual ones~\cite{arai2025numerical, yang2025robustness, satpute2024math, boye2025reasoning, rahman2025fragile}. To understand whether this limitation stems from how LLMs encode numerical information versus how they express it, we extracted embeddings from each transformer layer of fine-tuned models and trained supervised probes to predict material properties directly from these representations. This approach, inspired by probing classifiers used to study linguistic knowledge in NLP~\cite{skean2025layer, gupta2024depth}, reveals striking differences between properties---and identifies a fundamental bottleneck in LLM numerical prediction.

\begin{figure}[ht]
\centering
\includegraphics[width=\textwidth]{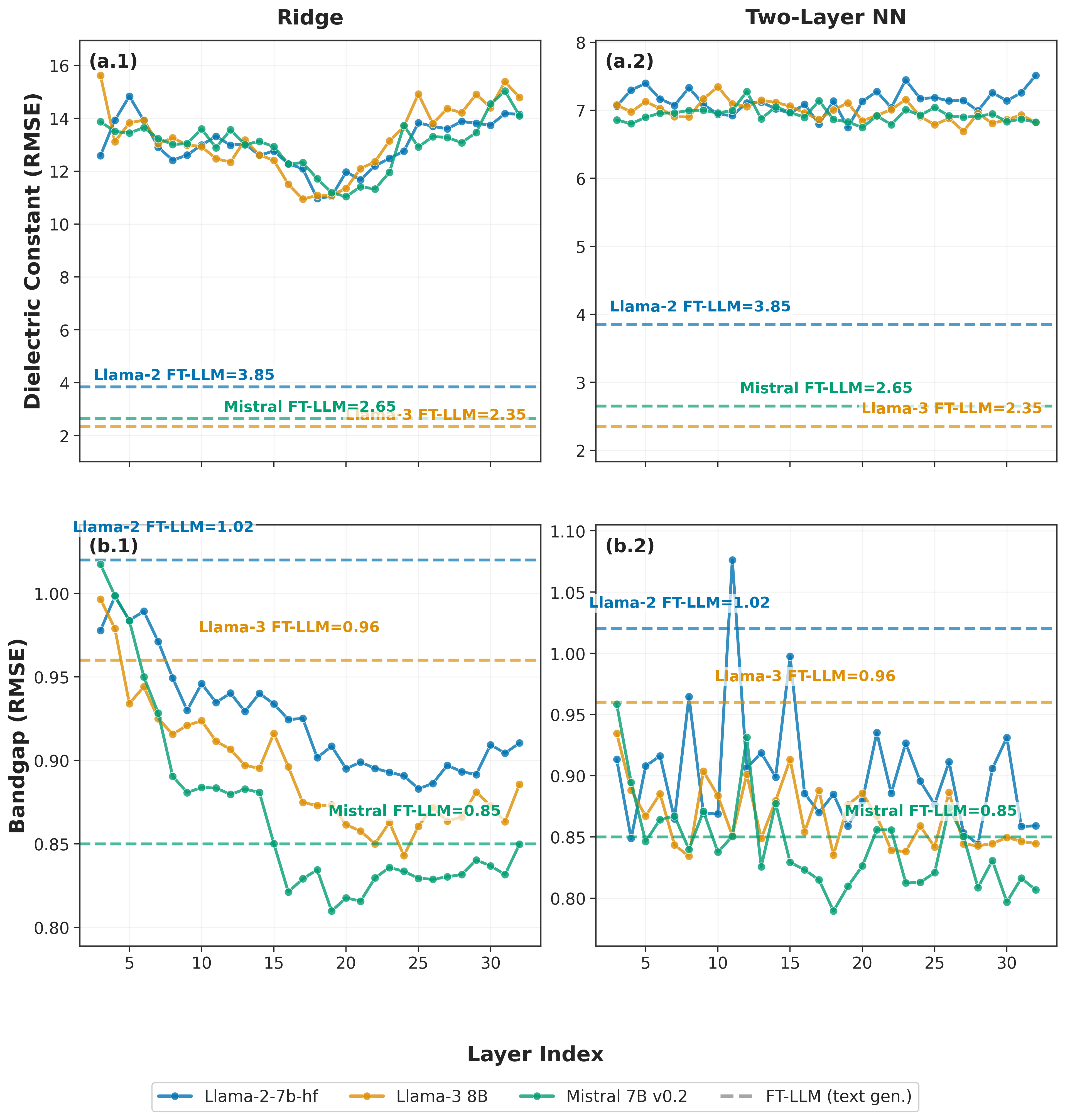}
\caption{\textbf{Layer-wise embedding probes for property prediction.} Test RMSE as a function of transformer layer for bandgap (top row) and dielectric constant (bottom row) across three model families, shown for ridge regression (left) and two-layer neural network (right). All three probe architectures (ridge, single-layer NN, two-layer NN) were tested with consistent results; the single-layer NN is omitted for clarity. Horizontal dashed lines indicate fine-tuned LLM text generation performance. For bandgap, intermediate layer embeddings match or exceed text output performance, suggesting a verbalization bottleneck. For dielectric constant, embeddings consistently underperform text generation by approximately 3$\times$, indicating property-dependent knowledge encoding.}
\label{fig:layerwise}
\end{figure}

For bandgap prediction, layer-wise probes achieve performance comparable to or better than the full fine-tuned model's text output (Figure~\ref{fig:layerwise}). Critically, this effect is consistent across all three model families tested: Llama-2-7B, Llama-3-8B, and Mistral-7B all show intermediate layer embeddings matching or exceeding the fine-tuned model's text generation RMSE. This pattern holds across all three probe architectures tested---ridge regression, single-layer neural network, and two-layer neural network (Figure~\ref{fig:layerwise} shows ridge and two-layer results)---ruling out model-specific artifacts. The implication is significant: bandgap information is encoded in the model's internal representations~\cite{chen2024states}, but the language modeling head provides no additional benefit---and may even degrade precision through the verbalization process. We term this the ``LLM head bottleneck'': fine-tuned models contain more predictive information in their hidden representations than they can express through autoregressive token generation~\cite{arai2025numerical}.

Dielectric constant prediction presents a starkly different picture. Despite identical experimental setup, embedding probes achieve RMSE of approximately 6--7, compared to fine-tuned text generation RMSE of 2.3---a persistent 3$\times$ performance gap. This gap remains consistent across all layers, probe architectures, and model families, indicating that dielectric knowledge is not accessible through standard embedding extraction. However, interpreting this result requires caution: the dielectric dataset contains only 4,870 samples, compared to 38,644 for bandgap and is heavily right-skewed, potentially limiting the probe's ability to learn robust mappings for high-value outliers.

We investigated several hypotheses to explain this asymmetry. The sample density hypothesis posited that dielectric's sparser training data (63 samples per unit of property range versus 1,164 for bandgap) might prevent effective embedding learning. To test this, we created a degraded bandgap dataset matching dielectric's sample density and distribution characteristics. Contrary to the hypothesis, embedding probes on degraded bandgap still achieved performance comparable to fine-tuned text output, ruling out sample density as the explanation (see S8 for details).

\begin{figure}[ht]
\centering
\includegraphics[width=\textwidth]{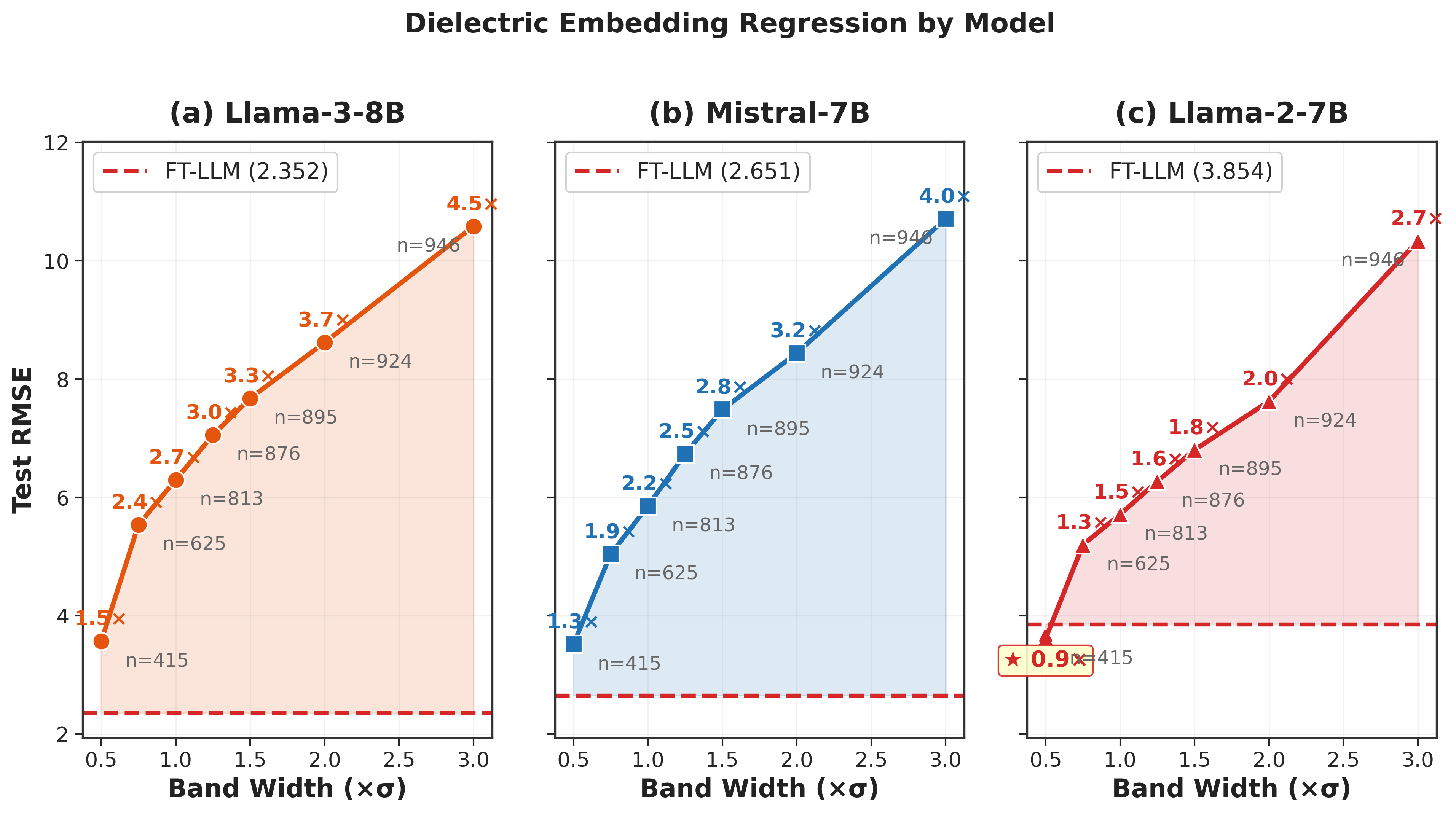}
\caption{\textbf{Dielectric prediction error varies with property range across all model families.} RMSE ratio (embedding probe / fine-tuned LLM) as a function of evaluation range for Llama-3-8B (left), Mistral-7B (center), and Llama-2-7B (right). All three models show consistent degradation as the evaluation range widens. Remarkably, Llama-2-7B embeddings outperform fine-tuned text generation at narrow band (0.9$\times$ FT-LLM at $\mu \pm 0.5\sigma$), while Mistral-7B achieves 1.3$\times$ and Llama-3-8B reaches 1.6$\times$. Performance degrades to 2.4--3.8$\times$ when including extreme values ($\mu \pm 3\sigma$). All models use Layer 17 embeddings and exhibit the same monotonic degradation pattern despite different FT-LLM baselines (2.4, 2.7, and 3.9 RMSE respectively).}
\label{fig:dielectric_range}
\end{figure}

Further analysis of dielectric prediction errors revealed a more nuanced picture that is consistent across all three model families tested (Figure~\ref{fig:dielectric_range}). When restricting evaluation to compounds with typical dielectric values (within $\pm 0.5\sigma$ of the mean), embedding probes achieve dramatically improved performance relative to the full-dataset gap. Remarkably, Llama-2-7B embeddings \textit{outperform} fine-tuned text generation at narrow band (0.9$\times$ FT-LLM), while Mistral-7B achieves 1.3$\times$ and Llama-3-8B reaches 1.6$\times$. Performance degrades monotonically as the evaluation range expands across all models: approximately 1.3--2.1$\times$ at $\pm 1\sigma$, 1.6--2.8$\times$ at $\pm 2\sigma$, and 2.4--3.8$\times$ at $\pm 3\sigma$. This consistency across model families indicates that embeddings do capture substantial dielectric information for typical materials but fail systematically for high-dielectric outliers---a property-dependent limitation rather than a model-specific artifact.

The contrast between bandgap and dielectric suggests property-dependent encoding mechanisms. For bandgap---a property extensively discussed in materials science literature and with well-established structure-property relationships~\cite{yang2025bandgap}---pre-trained LLMs appear to develop representations that encode predictive information accessible through linear probing. For dielectric constant---a tensor property with complex frequency dependence and less systematic coverage in training corpora~\cite{alampara2024mattext}---fine-tuning teaches knowledge that is not accessible through token embeddings alone. This interpretation aligns with recent work showing that LLMs store different types of knowledge in distinct architectural components~\cite{chen2024states, gwak2025layer}.

These findings have practical implications for materials informatics. For properties like bandgap, embedding extraction followed by lightweight regression may provide a computationally efficient alternative to full fine-tuning, particularly for screening applications where inference cost matters. For properties like dielectric constant, full fine-tuning remains necessary, and practitioners should be aware that high-value outliers may require additional attention or specialized approaches. The property-dependent nature of encoding also suggests caution when generalizing from benchmark performance on one property to expected performance on others---the mechanisms underlying prediction success may differ fundamentally.

\subsection{Temporal stability of API-based models}

\begin{figure}[ht]
\centering
\includegraphics[width=0.8\textwidth]{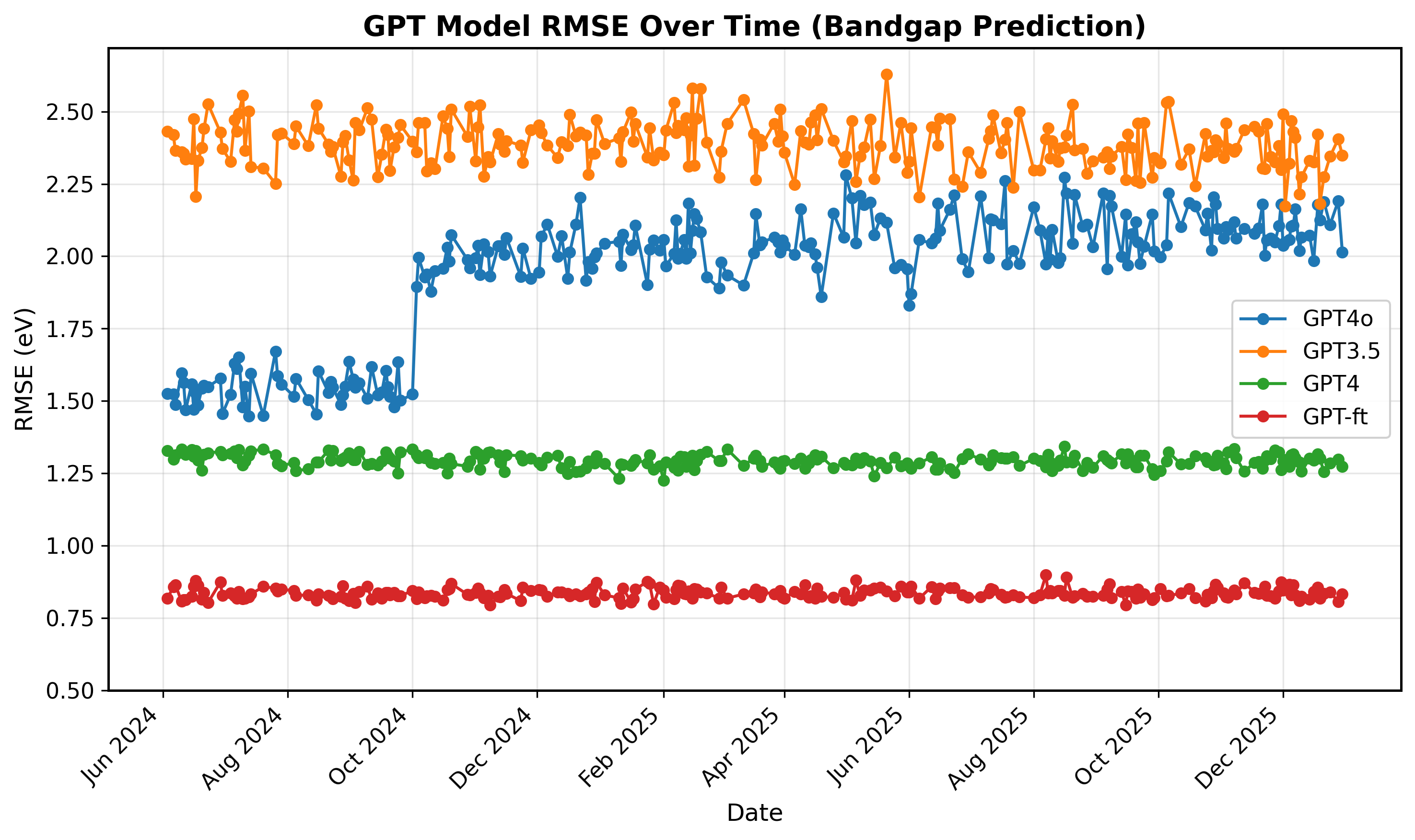}
\caption{\textbf{Temporal variation in GPT model performance.} RMSE for bandgap prediction across evaluation timepoints spanning 18 months. GPT-4o shows the largest variation (42.8\%), with a notable performance shift following the November 2024 endpoint update. Even fine-tuned models exhibit measurable drift over time.}
\label{fig:temporal}
\end{figure}

A critical but often overlooked concern for scientific applications of LLMs is reproducibility over time. Unlike downloadable open-weights models with fixed parameters, API-based models such as GPT present fundamental challenges for reproducible science. Users have no access to the internal architecture, weight configurations, or random seeds used during inference. This introduces built-in stochasticity that cannot be controlled or eliminated, even when using low temperature settings and identical prompts~\cite{wang2025robustness, arai2025numerical}. Additional concerns include silent endpoint updates, quantization changes, and model version swaps that may occur without explicit notification. To quantify these effects, we tracked GPT model performance on identical bandgap prediction tasks over an 18-month period (Figure~\ref{fig:temporal}).

\begin{table}[ht]
\centering
\caption{\textbf{GPT model performance variance over 18 months.} Maximum variation represents the largest percentage deviation from mean RMSE observed across evaluation timepoints. Data through December 2025.}
\label{tab:gpt_variance}
\begin{tabular}{lccc}
\toprule
Model & Mean RMSE (eV) & Max Variation (\%)\\
\midrule
GPT-FT & 0.84 & 13 \\
GPT-4 & 1.3 & 9 \\
GPT-4o & 1.9 & 43 \\
GPT-3.5 & 2.4 & 19 \\
\bottomrule
\end{tabular}
\end{table}

The results reveal substantial temporal instability across all GPT models (Table~\ref{tab:gpt_variance}). GPT-4 shows 9\% variation, GPT-3.5 shows 19\%, and even fine-tuned models (GPT-FT) exhibit 13\% variation, indicating that fine-tuning does not fully insulate against upstream changes or internal stochasticity. The most dramatic case is GPT-4o, where we observe a 43\% maximum variation. While this large value is attributable to an endpoint change on November 20, 2024~\cite{openai2024gpt4omodels}, the practical implication remains significant: a single undocumented API update caused model performance to shift by nearly half an RMSE unit over the course of one day. Even when the cause can be identified post-hoc, such sudden changes pose serious challenges for ongoing research projects, high-throughput screening campaigns, or any application requiring consistent predictions over time.

These findings have important implications for scientific reproducibility. Studies relying on API-based models should report exact model version strings and evaluation dates, as results may not be reproducible months or years later. For applications requiring long-term stability---such as high-throughput screening pipelines or regulatory submissions---open-weights models with frozen checkpoints may be preferable despite their higher computational requirements. The materials informatics community should consider developing standardized protocols for documenting and mitigating API model drift, analogous to software versioning practices in computational chemistry~\cite{dunn2020benchmarking}.


\section{Conclusion}\label{sec:conclusion}

This evaluation of 25 large language models across four materials science tasks reveals that output modality is the fundamental variable governing LLM behavior in scientific applications. Symbolic and numerical tasks produce qualitatively different failure modes with distinct implications for uncertainty quantification. Layer-wise probes uncover an LLM head bottleneck where intermediate representations encode more predictive information than autoregressive text generation can express, and knowledge graph analysis shows that fine-tuning builds distributional entity representations through co-occurrence rather than acquiring physical understanding. The poor performance of base models on knowledge graph tasks indicates that current LLMs capture far less domain-specific materials knowledge than structured extraction from scientific literature, and that this gap must be bridged through fine-tuning.

These findings suggest several directions for future investigation. The LLM head bottleneck implies that embedding extraction paired with lightweight supervised probes may offer a computationally efficient alternative to text generation for numerical property prediction, though systematically mapping which properties are amenable to this approach remains open. The distributional nature of knowledge graph performance raises a more fundamental question: can fine-tuned models generalize to unseen entities, or are they limited to recombining familiar patterns? More broadly, one implication of this study is that LLM performance on one task does not reliably predict performance on another, making multi-property, multi-task benchmarks necessary for holistic evaluation. Finally, confident hallucination — where models produce precise, consistent, and wrong numerical predictions — remains an unsolved reliability challenge that entropy-based uncertainty quantification cannot address; developing calibrated alternatives is a prerequisite for responsible deployment in materials discovery.


\section{Methods}\label{sec:methods}

\subsection{Datasets}

We evaluate LLM performance across four materials science tasks spanning regression, classification, and knowledge graph completion. The bandgap dataset comprises 38,644 compounds with DFT-computed band gaps ranging from 0 to 8 eV, sourced from the Materials Project~\cite{matbench2020}. The dielectric constant dataset comprises 4,870 compounds with static dielectric tensor traces ranging from approximately 1.5 to 100, also derived from Materials Project calculations. For crystal system classification, we use 45,089 compounds labeled with one of seven Bravais lattice systems (triclinic, monoclinic, orthorhombic, tetragonal, trigonal, hexagonal, cubic). The MatKG dataset~\cite{venugopal2024matkg} provides 5.4 million entity-relation-entity triples extracted from materials science literature, which we use for link prediction evaluation across 18 relation types. All datasets are split 80/20 for training and testing, with stratified sampling for classification tasks (see S1 for details).

\subsection{Models}

We benchmark 25 large language models spanning 10 architectural families. Open-weights models include Llama-2 (7B, 13B, 70B)~\cite{touvron2023llama2}, Llama-3 (8B, 70B)~\cite{meta2024llama3}, Mistral-7B and its instruction-tuned variants~\cite{jiang2023mistral}, Mixtral-8x7B~\cite{jiang2024mixtral}, CodeLlama (7B, 13B, 70B), Gemma (2B, 7B), Phi-3-mini, Zephyr-7B, and DeepSeek-R1-Distill-32B~\cite{deepseek2025r1}. Proprietary models include GPT-3.5-turbo, GPT-4o, GPT-4o-mini, GPT-4.1, and GPT-4.1-mini~\cite{openai2024gpt4}. This selection spans model sizes from 2B to over 200B parameters, enabling analysis of scaling effects on materials property prediction (see S2 for details).

\subsection{Fine-tuning}

All open-weights models are fine-tuned using Low-Rank Adaptation (LoRA)~\cite{hu2021lora, vanherck2025assessment, song2025csllm} applied to attention projection matrices (q\_proj, k\_proj, v\_proj, o\_proj). For models exceeding available GPU memory, we employ QLoRA~\cite{dettmers2023qlora} with 4-bit NF4 quantization. Fine-tuning proceeds for 10 epochs with learning rate $1 \times 10^{-5}$, batch size 1 with 16 gradient accumulation steps (effective batch size 16), Paged Adam optimizer, and 3\% warmup. OpenAI models are fine-tuned through their API with default hyperparameters. All fine-tuning infrastructure utilizes the Predibase~\cite{predibase2024} serverless platform (Ludwig 0.10.3) for open-weights models and the OpenAI fine-tuning API for proprietary models (see S4 for details).

\subsection{Inference}

For all evaluations, we use greedy decoding with temperature $T=0.1$ to balance determinism with output diversity. Each test sample is evaluated across 10 independent inference runs to compute prediction statistics. For regression tasks, we report the mode of parsed numerical outputs; for classification, we report the majority vote. Samples where the model fails to produce a parseable response across all 10 runs are excluded from metric computation. Prompt templates follow a consistent structure: system instruction specifying the task and the query compound formula (see S3 for prompt templates and details).

\subsection{Evaluation metrics}

Regression tasks (bandgap, dielectric constant) are evaluated using root mean squared error (RMSE). Crystal system classification uses accuracy. MatKG link prediction reports Top-1 and Top-5 accuracy, measuring whether the ground-truth entity appears in the model's top-ranked predictions.

\subsection{Response entropy}

To quantify prediction uncertainty, we compute response entropy~\cite{shannon1948mathematical} from the distribution of outputs across inference runs:
\begin{equation}
H = -\sum_{i} p_i \log p_i
\end{equation}
where $p_i$ is the frequency of unique output $i$ across runs. Low entropy indicates consistent predictions (high confidence), while high entropy suggests the model is uncertain. 

\subsection{Layer-wise embedding probes}

To investigate what materials knowledge is encoded in LLM representations, we extract hidden state embeddings from each transformer layer. Importantly, these embeddings are extracted from models with trained LoRA adapters loaded, not from base pretrained models, allowing us to probe what the fine-tuned models have learned. For each layer $l \in \{1, ..., L\}$, we extract the hidden state corresponding to the final token position after processing the input prompt. These embeddings are then used to train three types of supervised probes: (1) ridge regression with regularization $\alpha=1.0$, (2) a single hidden layer neural network with 256 units and ReLU activation, and (3) a two-layer neural network with 512 and 256 hidden units. Probes are trained on the same train/test splits as the main fine-tuning experiments. By comparing probe performance across layers, we identify where task-relevant information emerges in the model's forward pass, and by comparing probe performance to full fine-tuned model performance, we assess how much additional value the language modeling head provides over direct regression on embeddings.

\subsection{Cross-task transfer}

To evaluate whether fine-tuning on one materials task transfers to others, we construct a $4 \times 4$ transfer matrix. Each model fine-tuned on task $A$ is evaluated on all tasks $\{A, B, C, D\}$, comparing performance to the base (non-fine-tuned) model. Statistical significance is assessed by comparing the performance difference to the base model's variance across inference runs: improvements or degradations exceeding one standard deviation of base model performance are marked as statistically significant. This enables analysis of which task combinations exhibit positive transfer, negative interference, or negligible effect.

\subsection{Temporal stability}

To assess reproducibility of API-based models over time, we conduct identical evaluations of GPT models at multiple time points spanning 18 months. Using frozen prompt templates and identical test sets, we track changes in RMSE, accuracy, and output distributions. This enables detection of performance drift due to model updates, quantization changes, or infrastructure modifications that may occur without explicit versioning by API providers~\cite{openai2024gpt4omodels}.

\subsection{Reproducibility}

All code and notebooks are available at \url{https://github.com/olivettigroup/LLM-Probe}. Datasets and Fine-tuned LoRA adapters for Llama-2, Llama-3, Mistral, and Mixtral are also available on Hugging Face under the \texttt{vinven7} namespace. Random seeds are fixed across all experiments. For API-based models, we record model version strings and API response metadata. Fine-tuning hyperparameters, hardware specifications, and training logs are provided in the supplementary materials.


\bibliography{references}

\end{document}